# Spin-imbalance in a one-dimensional Fermi gas

Yean-an Liao[1*], Ann Sophie C. Rittner[1*], Tobias Paprotta[1], Wenhui Li[1,†], Guthrie B. Partridge[1,‡], Randall G. Hulet[1], Stefan K. Baur[2] & Erich J. Mueller[2]

*[1]Department of Physics and Astronomy and Rice Quantum Institute, Rice University, Houston, TX 77251, USA.*

*[2]Laboratory of Atomic and Solid State Physics, Cornell University, Ithaca, NY 14853, USA.*

*[\*] These authors contributed equally to this work.*

[†] Permanent address: Centre for Quantum Technologies, National University of Singapore, 3 Science Drive 2, Singapore, 117543.

[‡] Present address: Laboratoire Charles Fabry de l'Institut d'Optique, UMR CNRS 8501, Palaiseau, France.

**Superconductivity and magnetism generally do not coexist. Changing the relative number of up and down spin electrons disrupts the basic mechanism of superconductivity, where atoms of opposite momentum and spin form Cooper pairs. Nearly forty years ago Fulde and Ferrell[1] and Larkin and Ovchinnikov[2] (FFLO) proposed an exotic pairing mechanism where magnetism is accommodated by formation of pairs with finite momentum. Despite intense theoretical and experimental efforts, however, polarised superconductivity remains largely elusive[3]. Here we report experimental measurements of density profiles of a two spin mixture of ultracold [6]Li atoms trapped in an array of one dimensional (1D) tubes, a system analogous to electrons in 1D wires. At finite spin imbalance, the system phase separates with an inverted phase profile as compared to the three-dimensional (3D) case. In 1D, we find a partially polarised core surrounded by wings composed of**

**either a completely paired or a fully polarised Fermi gas, depending on the degree of polarisation. At zero temperature, this system is predicted to be a 1D analogue of the FFLO state[4-12]. This study demonstrates how ultracold atomic gases in 1D may be used to create non-trivial new phases of matter, and also paves the way for direct observation and characterization of the FFLO phase.**

The FFLO states are perhaps the most interesting of a number of exotic polarised superconducting phases proposed in the past 40 years. In the original concept of Fulde and Ferrell (FF), Cooper pairs form with finite centre of mass momentum[1]. Larkin and Ovchinnikov (LO) proposed a related model where the superconducting order parameter oscillates in space[2]. These two ideas are closely related, as the oscillating order parameter may be interpreted as an interference pattern between condensates with opposite centre of mass momenta. The spin density oscillates in the LO model, leading to a build-up of polarisation in the nodes of the superconducting order parameter. Thus the LO state can be considered a form of micro-scale phase separation with alternating superfluid and polarised normal regions. By including more and more momenta, subsequent theorists were able to evaluate the stability of ever more complicated spatial structures[3].

Previous studies of superfluidity in fermionic atoms show that ultracold atoms are a powerful tool for investigating the emergent properties of interacting systems of many particles. While largely analogous to an electronic superconductor, the atomic systems feature tunable interactions. This extra degree of control has lead to a number of unique experiments and conceptual advances. Furthermore, the absence of spin relaxation enables us to spin-polarise the atoms in order to explore the interplay between magnetism and superfluidity, with the potential to observe the FFLO phase. Recent calculations indicate that if a FFLO phase exists in three dimensional trapped gases, it will occupy a very small volume in parameter space[13,14]. Experiments in 3D and in the strongly-interacting limit show that the gas phase separates with an unpolarised superfluid core surrounded by a polarised shell[15-19], with no evidence for the FFLO phase. Here, we study a polarised Fermi gas in 1D, where theory predicts that a large fraction of the phase diagram is occupied by an FFLO-like phase (see Fig. 1a)[4-12]. In this 1D setting, the physics should be closest to that described by LO, where an oscillating superfluid order parameter coexists with a spin-density wave. Due to fluctuations, the order will be algebraic rather than long-ranged. The increased stability of FFLO-like phases in 1D can be understood as a "nesting" effect, where a single wavevector connects all points on the Fermi surface, allowing all atoms on the Fermi surface to participate in finite momentum pairing, while in 3D, only a small fraction of these atoms are able to do so. Similar enhancements are predicted for systems of lattice fermions and quasi-1D geometries[10,20].

Our work complements studies of astrophysical objects[3] and solid state systems. Like our current experiment, the solid state experiments typically involve highly anisotropic materials – made up either of weakly coupled two-dimensional planes or 1D wires. Examples include the organic superconductor $\lambda$-(BETS)$_2$FeCl$_4$ [21] and the heavy fermion superconductor CeCoIn$_5$ [22,23]. However, FFLO states have not been conclusively observed in any system.

Details of our experimental procedures are given in Methods and in Refs. 16 and 17. We create a mixture of the two lowest hyperfine levels of the $^6$Li ground state, the majority state $|1\rangle$, and the minority state $|2\rangle$. An array of 1D tubes is formed with a 2D optical lattice[24]. The lattice potential is given by $V = V_0 \cos^2(kx) + V_0 \cos^2(ky)$, with $k = 2\pi/\lambda$ and $V_0 = 12\ \varepsilon_r$, where $x$ and $y$ are two orthogonal radial coordinates, $\lambda$ is the optical trap laser wavelength of 1064 nm, and $\varepsilon_r = \hbar^2 k^2/2m$ is the recoil energy. There are several requirements for the system to be 1D. First, only the lowest transverse mode in each tube may be populated. This requires that both the thermal energy $k_B T$ and the 1D Fermi energy $\varepsilon_F = N_1 \hbar \omega_z$ be small compared to the transverse confinement energy $\hbar \omega_\perp$. Here $N_1$ is the number of atoms per 1D tube in state $|1\rangle$, and $\omega_z$ and $\omega_\perp$ are the axial and transverse confinement frequencies of an individual tube. Second, the single particle tunnelling rate $t$ should be small compared to the larger of $\varepsilon_F$ and $T$ [7,10] The condition $\varepsilon_F > t$ is equivalent to specifying that the Fermi surface is one dimensional, while the condition $T > t$ makes the inter-tube coupling incoherent. All conditions are well satisfied in our experiment: the tube aspect ratio $\omega_\perp / \omega_z = 1000$ is larger than $N_1 \approx 120$ for the central tube; and $t/k_B \approx 17$ nK is much smaller than both $\varepsilon_F / k_B \approx 1.2$ $\mu$K, and $T \approx 175$ nK.

We tune an external magnetic field to the BCS side (890 G) of the broad 3D Feshbach resonance in $^6$Li[25,26], where the 1D interactions are strongly attractive[27,28]. We measure the *in situ* density of the two spin species by sequential imaging with two probe laser beams, choosing their intensity and frequency to maximize the signal to noise ratio of the density difference (see Methods). Assuming hydrostatic equilibrium, the 1D spatial density profiles $n_{1,2}(z)$ can be expressed in terms of $\mu = \mu_0 - V(z)$, and $h = h_0$, where $\mu_0$ and $h_0$ are the chemical potential and chemical potential difference at the centre of the tube, set by the total number of particles in the tube $N = N_1 + N_2$ and polarisation $P = (N_1-N_2)/N$; $V(z)$ is the axial confinement potential. In particular, the phase boundary between the fully paired and partially polarised regions occurs where the density difference $n_1(z) - n_2(z) = 0$, and the boundary between the fully and partially polarised phases corresponds to $n_2(z) = 0$, as shown in Fig. 1b.

Figure 2 shows axial density profiles of state $|1\rangle$, state $|2\rangle$, and their difference for a range of polarisations. These images represent the sum of the linear density in all tubes in

our system, and are produced by integrating our column density images across the remaining transverse direction. At low polarisation, a partially polarised region forms at the centre of the trap (Fig. 2a) whose radius increases with increasing polarisation (Fig. 2b). This is distinctly different from a polarised 3D gas where the centre is fully paired. At a critical polarisation, $P_c$, the partially polarised region extends to the edge of the cloud (Fig. 2c). When the polarisation increases further, the edge of the cloud becomes fully polarised (Fig. 2d). From the images of the atomic clouds we extract the axial radii of the ensemble of tubes of the minority density and the density difference. The axial radii of the tube bundle are equivalent to the central tube radius for our experiment because the inner and outer boundaries both decrease monotonically going from the central to the outer tubes (see supplementary material). We perform an inverse Abel transform to obtain the number of particles and polarisation in the central tube. Following Ref. 6, we plot these radii as a function of the central tube polarisation (Fig. 3), normalizing the radii by $(N_0)^{1/2} a_z$ where $N_0$ is the total number in the central tube and $a_z = (\hbar/m\omega_z)^{1/2}$ is the harmonic oscillator length along the central tube. The critical polarisation $P_c$ corresponds to the crossing of these two radii where the entire cloud is partially polarised: for $P < P_c$ the radius of the density difference is smaller than the minority radius, while the opposite occurs for $P > P_c$. From a linear fit to the data, we find $P_c = 0.13$ +/- 0.03. We use the thermodynamic Bethe ansatz to calculate theoretical density distributions, and carry out an identical analysis. We find quantitative agreement with the experimental density profiles, with a best fit temperature of $T = 175$ +/- 50 nK = 0.15 $T_F$ (see supplementary material). The theoretical density profiles yield $P_c = 0.17$ with weak temperature dependence.

Although the strength of the FFLO correlations fall off with temperature, theoretical calculations[29] predict that at $T = 0.15$ $T_F$ there are a range of polarizations near $P_c$ for which they remain at detectable levels. Given that our interactions are stronger than those studied in Ref. 29 and that stronger interactions make the low temperature phases more robust[11], we expect that significant FFLO-like correlations should be present in our system for a range of polarizations.

We have created a strongly interacting, two-component Fermi gas in 1D and measured its phase diagram as a function of polarization. The system is at sufficiently low temperature to observe three distinct phases in agreement with theory. This is an example of an optical lattice-based quantum simulator that produces a phase diagram of non-trivial quantum phases. While we have not directly observed the FFLO phase, the observed density profiles agree quantitatively with theories that exhibit the 1D equivalent of FFLO correlations at low T[11,29]. In the future, we intend to measure the pair momentum distribution of the partially polarised phase in order to directly reveal its FFLO correlations.

## Methods Summary

We start from quantum degenerate, spin-imbalanced $^6$Li Fermi gas in a single beam far-off resonance optical trap[16,17], which is then loaded into a crossed beam optical dipole trap formed by a pair of retro-reflected beams propagating in the $x$ and $y$ directions. We turn on the 2D lattice by ramping up the optical trap laser power and rotating the polarisations of the retro-reflected beams to create standing waves in two orthogonal directions. The intersection of the standing waves creates 1D tubes with an energy depth of 12 $\varepsilon_r$ in the central tube, with $\omega_{\perp, z} = (2\pi)$ $(2{\times}10^5, 200)$ Hz. At a global polarisation $P \approx 0$, the total number of atoms is ~4×10$^5$, giving a total number of atoms in the central tube of $N$ ~240 ± 20. The column densities of each state and their difference is obtained from two *in situ* polarisation phase contrast images[30] taken in rapid succession and with different detuning. The temperature is determined by fitting the *in-situ* density of a balanced spin mixture to a Thomas-Fermi distribution and is measured to be $T < 0.05\ T_F$ before turning on the 2D lattice and $T \sim 0.09 \pm 0.03\ T_F$ after slowly turning on the lattice and then, slowly rotating the polarisation back to the 3D trap configuration. The temperature in the lattice is estimated from the *in situ* density distributions.

**Full Methods** and any associated references are available in the online version of the paper at www.nature.com/nature.

**Supplementary Information** accompanies the paper on **www.nature.com/nature**

**Acknowledgements** We thank S.E. Pollack for providing software to remove fringes from the images and Melissa Revelle for help on the experiment. EJM would like to thank Carlos Bolech and Paata Kakashvili for discussion of techniques for analysing the data. Supported under ARO Award W911NF-07-1-0464 with funds from the DARPA OLE Program, and by the NSF, ONR, and the Welch (grant C-1133) and Keck Foundations.

**Author's Contributions**

Y.A.L., T.P., A.S.R., G.B.P., W.L. and R.G.H. constructed the apparatus. Y.A.L., A.S.R., and T.P. acquired and processed the data. S.K.B. and E.J.M. did the theory and extracted the phase boundaries from the data. R.G.H and E.J.M. supervised the investigation. All contributed to writing the manuscript.

**Author Information** Reprints and permissions information is available at

npg.nature.com/reprintsandpermissions. The authors declare no competing financial interests. Correspondence and requests for materials should be addressed to Randall G. Hulet (randy@rice.edu).

## Methods

**Preparation.** We produce a quantum degenerate, strongly interacting, spin-imbalanced $^6$Li Fermi gas using our previously published methods[16,17]. Starting from a quantum degenerate gas of $^6$Li in a single beam far-off resonance optical dipole trap, we control the relative population of two hyperfine states, $F = \frac{1}{2}$, $m_F = \frac{1}{2}$ (state $|1\rangle$) and $F = \frac{1}{2}$, $m_F = -\frac{1}{2}$ (state $|2\rangle$), where $F$ is the total spin and $m_F$ is the projection along the quantization axis, by driving radio frequency sweeps between them at different power. The spin mixture is created in a uniform magnetic field at 765 G within the broad Feshbach resonance between states $|1\rangle$ and $|2\rangle$ centred at 834 G[25,26]. Atoms are evaporatively cooled by lowering the trap depth in the single beam optical trap. During evaporation, the field is adiabatically swept to 890 G, on the BCS side of the Feshbach resonance where the 3D scattering length $a_{3D}$ = - 9145 $a_o$ ($a_o$ is the Bohr radius). At the end of evaporation, we turn on a crossed-beam optical dipole trap formed by two orthogonal, retro-reflected laser beams, with elliptical laser beam waists ($1/e^2$ radii) of 54 µm by 236 µm, with the beams propagating in the $x$-$y$ plane and the long axes of the ellipses oriented along $z$. The polarisation of each retro-reflected beam is controlled by liquid crystal variable retarders (LCVRs) and is perpendicular to that of the incident beam in the trap configuration. The trap depth is 0.5 μK with axial and the radial trapping frequencies of 50 and 153 Hz, respectively. We then turn on the optical lattice by simultaneously ramping up the laser power and rotating the polarisation of each retro beam to be parallel to its corresponding incident beam, resulting in a 2D lattice of 1D tubes. The lattice turn-on time constants are 130 ms for intensity and 70 ms for polarisation, with both having smooth error-function like trajectories, optimised to minimise heating. The final 1D lattice depth is 12 $\varepsilon_r$ ($\varepsilon_r$ = 1.39 μK) with radial and axial trapping frequencies in the central tube ($\omega_\perp$, $\omega_z$) = ($2\pi$) ($2\times 10^5$, 200) Hz. After waiting 50 ms, we take images that record the column densities of each state in the array of 1D tubes. Under these conditions $a_\perp$ = 1720 $a_o$, $a_z$ = 5.3 × $10^4$ $a_o$, and $a_{1D}$ = 2099 $a_o$, where $a_{\perp,z}$ = ($\hbar/m\omega_{\perp,z}$)$^{1/2}$ and $a_{1D}$ is the 1D scattering length defined in Ref. 31.

**Imaging**. The column densities of each state and their difference is obtained from two *in situ* phase-contrast polarisation imaging (PCPI)[30] shots, taken in rapid succession and with different detunings near the $^2S_{1/2}$ to $^2P_{3/2}$ atomic transition. Imaging 1D gases *in situ* is problematic due to high optical densities and heating from the first laser pulse. The first pulse dissociates atom pairs and the release of binding-energy affects the second image. At 890 G, the binding energy in 1D is ~ 6 μK, while in 3D this field corresponds to the BCS limit where there is little pairing energy. To minimise heating effects in the second image we use PCPI imaging with short intervals (as short as 5 μs) between images (See supplementary information for more imaging details).

**Temperature**. In the absence of the optical lattice an effective temperature is measured by fitting finite-temperature Thomas-Fermi distributions to clouds prepared with $P$ = $0$[16,32]. Before turning on the lattice, the effective temperature is <0.05 $T_F$ in the shallow trap, where $T_F$ is the Fermi temperature of a non-interacting gas of $N_1$ fermions[16]. In the lattice, temperature is measured by comparing the experimental column densities with the theory described in the next section.

**Description of the spin imbalanced attractive 1D gas.** Sufficiently far on the attractive side of the confinement induced resonance (CIR), the one dimensional spin-imbalanced attractively interacting Fermi gas may be described by the exactly solvable Gaudin-Yang model[33-35] with Hamiltonian

$$H = \sum_{\sigma=1,2} \sum_{i=1}^{N_\sigma} \frac{p_{i\sigma}^2}{2m} + g_{1D} \sum_{i=1}^{N_1} \sum_{j=1}^{N_2} \delta(z_{i1} - z_{j2}),$$

where the 1D coupling constant is related to the 1D scattering length by $g_{1D} = -\dfrac{2\hbar^2}{ma_{1D}}$.

An energy scale is given by the binding energy of the contact interaction $\varepsilon = \dfrac{\hbar^2}{ma_{1D}^2}$. To obtain the equation of state at a finite temperature $t = k_B T / \varepsilon$ (from now on we put $\varepsilon = 1$) we numerically solve a truncated set of the thermodynamic Bethe ansatz (TBA) equations[12,34], given by two nonlinear integral equations

$$\varepsilon(k) = 2(k^2 - \mu - \frac{E_B}{4\varepsilon}) + \frac{t}{\pi} \int_{-\infty}^{\infty} \frac{\mathrm{d}q}{1+(k-q)^2} \ln(1 + e^{-\varepsilon(q)/t}) + \frac{2t}{\pi} \int_{-\infty}^{\infty} \frac{\mathrm{d}q}{1+4(k-q)^2} \ln(1 + e^{-\kappa(q)/t})$$

$$\kappa(k) = k^2 - \mu - h + \frac{2t}{\pi} \int_{-\infty}^{\infty} \frac{\mathrm{d}q}{1+4(k-q)^2} \ln(1 + e^{-\varepsilon(q)/t})$$

where $\mu = \dfrac{\mu_1 + \mu_2}{2}$, $h = \dfrac{\mu_1 - \mu_2}{2}$. We have modified the original equations by replacing the binding energy of the contact interaction with the true two body binding energy $E_B$ in a harmonic waveguide in order to have the proper definition for the chemical potentials. Densities $n_{1,2}$ are obtained from the solution of two coupled linear integral equations[35] (similar to the equations for the zero-temperature Bethe ansatz)

$$(1 + e^{\varepsilon(k)/t})\sigma(k) = \frac{1}{\pi} - \frac{1}{\pi}\int\limits_{-\infty}^{\infty}\frac{dq\,\sigma(q)}{1+(k-q)^2} - \frac{2}{\pi}\int\limits_{-\infty}^{\infty}\frac{dq\,\rho(q)}{1+4(k-q)^2}$$

$$(1 + e^{\kappa(k)/t})\rho(k) = \frac{1}{2\pi} - \frac{2}{\pi}\int\limits_{-\infty}^{\infty}\frac{dq\,\sigma(q)}{1+4(k-q)^2}$$

as

$$n_2 a_{1D} = 2\int\limits_{-\infty}^{\infty}dk\,\sigma(k)$$

$$(n_1 - n_2)a_{1D} = 2\int\limits_{-\infty}^{\infty}dk\,\rho(k)\,.$$

This truncation is accurate when thermal fluctuations are unable to break the tightly bound pairs, i.e. when $k_B T/\varepsilon \ll 1$ (for a detailed discussion of a similar approximation see Ref. 37). In the experiment, $k_B T/\varepsilon \approx 0.02\text{-}0.03$, and we have explicitly checked higher order terms in the TBA equations, and have seen that they are small. Confinement effects[36,38,39] can modify the interactions between pairs and excess fermions in a way not captured by the Gaudin-Yang model. A study of the three and four-body problem carried out in Ref. 40 shows that for our experimental parameters, $a_\perp/a_{3D} \sim$ -0.19, these confinement effects shift energies by ~10%.

At strong coupling ($n^{1D}a_{1D} \to 0$) the equation of state of the Gaudin-Yang model reduces to that of a Tonks gas of bosons and a free Fermi gas[6,7]. This simplicity hides the fact that there are FFLO correlations in the system, with the many-body wavefunction changing sign whenever a boson crosses a fermion (see Supplementary Information).

**Calculation of density profiles.** For each tube, we use the Thomas-Fermi local density approximation to calculate the 1D density profiles, but allow the chemical potential to vary arbitrarily from one tube to the next

$$n_\sigma^{1D}(\rho, z) = n_\sigma^{1D}(\mu^c(\rho) - \tfrac{1}{2} m \omega_z^2 z^2, h(\rho), t),$$

where $\sigma = 1, 2$. $\mu^c(\rho), h(\rho)$ are related to the particle numbers for a tube a distance $\rho$ from the central axis by

$$N_\sigma(\rho) = \int_{-\infty}^{\infty} dz \, n_\sigma^{1D}(\mu^c(\rho) - \tfrac{1}{2} m \omega_z^2 z^2, h(\rho), t).$$

Numerically inverting this equation, we find the central chemical potentials $\mu^c(\rho), h(\rho)$. $N_\sigma(\rho)$ is obtained from the experimental data by inverse Abel transforming the radial profiles

$$n_{r,\sigma}(y) = \int_{-\infty}^{\infty} dz \, n_{c,\sigma}(y, z) = \frac{4}{\lambda^2} \int_{-\infty}^{\infty} dx \, N_\sigma(\sqrt{x^2 + y^2}) \text{ through}$$

$$N_\sigma(\rho) = \frac{-\lambda^2}{4\pi} \int_{\rho}^{\infty} dy \, \frac{\partial_y n_{r,\sigma}(y)}{\sqrt{y^2 - \rho^2}},$$

where $n_{c,\sigma}(y, z) = \frac{4}{\lambda^2} \int_{-\infty}^{\infty} dx \, n_\sigma^{1D}(\sqrt{x^2 + y^2}, z)$ is the column density. We fit the radial densities $n_{r,\sigma}(y)$ to a simple functional form, and analytically perform the integrals. We use the extracted $N_\sigma(\rho)$ to normalize our radii in Fig. 3.

**Figure 1: Theoretical *T* = 0 phase diagram (adapted from Ref. 6).**

**a,** Schematic with $\mu = \frac{1}{2}(\mu_1 + \mu_2)$ vs. $h = \frac{1}{2}(\mu_1 - \mu_2)$ showing three phases: fully paired (green), fully polarised (blue), and partially polarised (yellow) that is predicted to be FFLO. In a trap, $\mu$ decreases from the centre to the edge, while $h$ is constant throughout the tube. The vertical arrows show two possible paths from trap centre to edge: The partially polarised centre is surrounded either by a fully paired superfluid phase at low $h$ or by a fully polarised phase at high $h$. At a critical value of $h$, corresponding to a polarisation $P_c$, the whole cloud is partially polarised. **b,** Phase diagram of the 1D trapped gas with infinitely strong point interactions. The scaled axial radius is defined in the Fig. 3 caption. The red line corresponds to the scaled radius of the density difference, while the blue line is the scaled radius of state $|2\rangle$.

**Figure 2: Axial density profiles of a spin-imbalanced 1D ensemble of tubes.** Integrated axial density profiles of the tube bundles (black circles: majority; blue diamonds: minority; red squares: difference) are shown as function of central *P*. (**a-d**) *P* corresponds to 0.015, 0.055, 0.10, and 0.33, respectively. **a,** At low *P*, the edge of the cloud is fully paired and the density difference is zero. The centre of the cloud is partially polarised. The density difference has been multiplied by 2 for better visibility of the phase boundary (dashed black line). **b,** For increasing *P*, the phase boundary moves to the edge of the cloud as the partially polarised region grows. **c,** Near $P_c$, where nearly the entire cloud is partially polarised. **d,** Well above $P_c$, where the edge of the cloud is fully polarised and the minority density vanishes.

**Figure 3: Experimental phase diagram as a function of polarisation in the central tube.** The scaled radii of the axial density difference (red diamonds) and the minority state ($|2\rangle$) axial density (blue circles) compared with a 175 nK Bethe ansatz calculation (solid lines). The dimensionless scaled axial radius, $R$ / ($a_z$ $N_0^{1/2}$), is plotted, where $R$ is the position along the bundle of tubes where the respective density vanishes, $N_0$ is total number in the central tube, and $a_z$ is the axial harmonic confinement length. At $P \sim 0.13 \pm 0.03$, both radii intersect indicating that the entire cloud is partially polarised. The data is in reasonable agreement with the theoretical crossing at slightly higher polarisation $P \sim 0.17$.

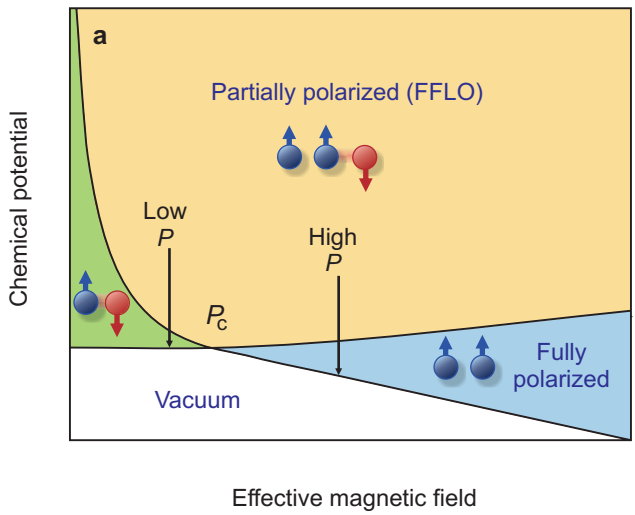

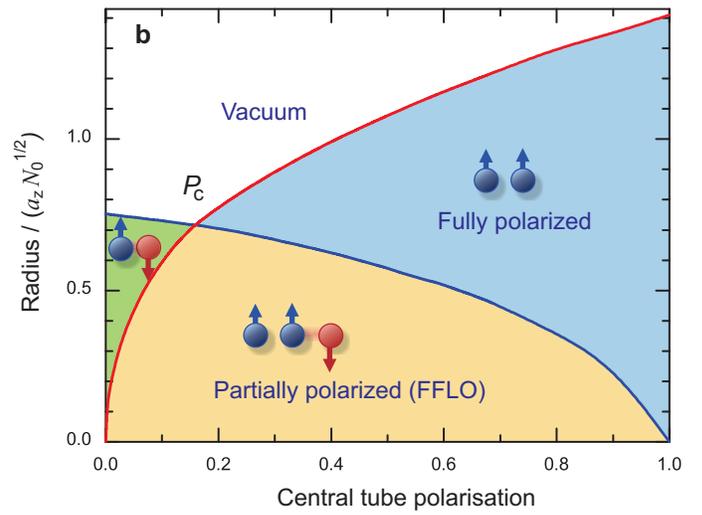

Author: Liao et. al.
Figure 1

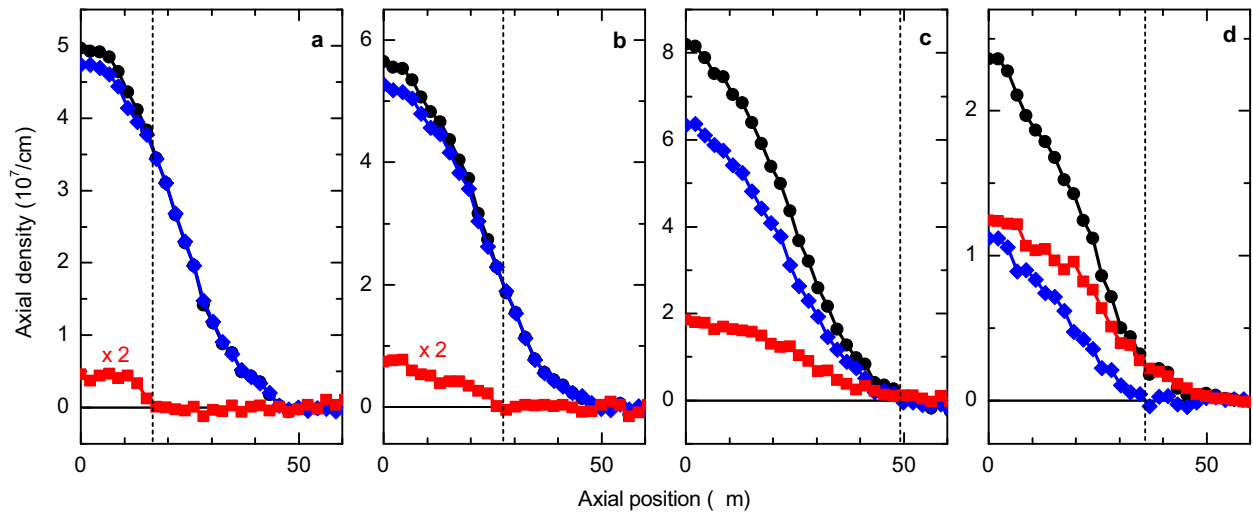

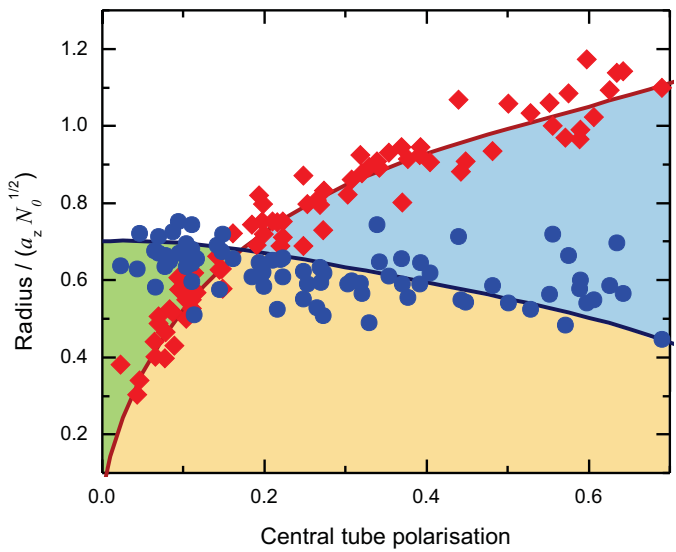

**Supplementary information**

**Imaging.** We use polarisation phase contrast imaging, as illustrated in Fig. S1. The first probe pulse is detuned to the red of the state $|2\rangle$ transition by 15 $\Gamma$, where $\Gamma$ = 5.9 MHz is the linewidth of the probing transition. This corresponds to a red detuning from the state $|1\rangle$ transition of 28.3 $\Gamma$ since states $|1\rangle$ and $|2\rangle$ are separated by 78.5 MHz (see Fig. S1b). The second laser pulse is tuned in between the state $|1\rangle$ and $|2\rangle$ transitions. The intensity of the first probe pulse is lower to minimize heating, while the intensity of the second is higher to improve the signal-to-noise ratio. Both phase contrast images are linear combinations of the two states with different weighting factors: the first image involves a sum, while the second shot is the difference (an example is shown in Fig. S1c). The two images are deconvolved to obtain column density images of $|1\rangle$ and $|2\rangle$, which we subsequently integrate along the remaining radial direction to obtain axial densities. The imaging resolution of ~3 μm is insufficient to resolve individual 1D tubes separated by $\lambda/2$ ~ 0.5 μm.

**Temperature.** As described in the methods section, we determine the temperature by comparing the experimental column densities to those expected from our thermodynamic Bethe ansatz calculation. Figure S2 shows the temperatures found for the 95 data sets used in Fig. 3. In the inset we illustrate the quality of the fit by plotting the reduced chi squared as a function of temperature for a representative data set,

$$\chi^2 = \frac{1}{2N}\sum_i \left[ \left(\frac{n_{\exp}^{\uparrow}(i) - n_{th}^{\uparrow}(i,T)}{\sigma_{\uparrow}}\right)^2 + \left(\frac{n_{\exp}^{\downarrow}(i) - n_{th}^{\downarrow}(i,T)}{\sigma_{\downarrow}}\right)^2 \right]$$ The sum is over each of the

$N = 65 \times 41 = 2665$ pixels. The experimental column density is $n_{\exp}^{\uparrow,\downarrow}(i)$ while the theoretical one with temperature $T$ is $n_{th}^{\uparrow,\downarrow}(i,T)$. This $\chi^2$ corresponds to the negative log likelihood distribution under the assumption that the noise is gaussian and independent from pixel to pixel. We estimate $\sigma_{\uparrow}$ and $\sigma_{\downarrow}$ as the standard deviation of the noise in a region of the column density image where no atoms are present. The statistical uncertainties in the temperature of the fit are far smaller than the systematic uncertainty in the axial trapping frequency. The 5% uncertainty in the axial trapping frequency maps onto a similar uncertainty in temperature.

**Central tube radius.** The axial radii in Fig. 3 are extracted from the axial densities, which are an average over the entire bundle of tubes. They represent the largest radius beyond which the minority or difference density is zero for *all* of the tubes. Here we empirically show that the radii for which these densities vanish is always greatest in the central tube, and hence the radii in Fig. 3 may be interpreted as a property of the central tube.

To demonstrate this equivalence we axially integrate the column densities and apply an inverse Abel transform to extract the number of up and down spin particles in each tube as function of radius (see inset of Fig. S3). From the number of particles and polarisation in a given tube we use the zero temperature Bethe ansatz to predict the axial position, $z$, at which the minority or difference density should vanish in that tube. Figure S3 shows the axial radii of minority (blue) and difference (red) as function of radial position, $\rho$, for a representative high and low polarization data set. Both phase boundaries are maximised at the central tube and decrease monotonically as one approaches the outer tubes. This monotonicity allows us to extract the central tube phase boundaries from the integrated column density profiles.

**Truncation to 1D model and the 1D FFLO-Bose/Fermi mixture crossover.** Here we give a brief overview of the physics that we expect to emerge as the interactions between the trapped atoms are made stronger, by either reducing the magnetic field or the density. Not only will this discussion help identify interesting directions of future research, but it helps to understand the limits of validity of our approximations. Our central observation is that our current experiments are accurately described by a model of 1D fermions, which are theoretically expected to display FFLO correlations, but by making small changes to the experiment we can enter a regime where the system behaves in a qualitatively different manner.

In 1998, Maxim Olshanii[1] addressed the problem of atoms interacting with 3D scattering lengths $a_{3D}$ trapped in an elongated tube with harmonic transverse confinement $V_\perp = m\omega_\perp^2 r^2 / 2$. By solving the two-body problem Olshanii showed that the low energy scattering properties between pairs of atoms are completely described by an effective 1D point interaction, whose strength is parameterized by a 1D scattering length, $a_{1D} = -\dfrac{a_\perp^2}{a_{3D}}\left(1 - C\dfrac{a_{3D}}{a_\perp}\right)$, where $a_\perp = \sqrt{\hbar/m\omega_\perp}$, and C ~ 1.033. The effective interactions switch from attractive to repulsive at the confinement induced resonance, $a_{3D} = a_\perp / C$. Our experiments are well on the attractive side of the resonance, with $a_\perp/a_{3D}$ = -0.18. Later, with Bergeman and Moore[2], Olshanii showed that as one approaches the confinement induced resonance, the energy of the two-body bound state begins to deviate from what one would expect from a strictly 1D model of fermions with point interactions. This deviation is understood by noting that when the interactions become strong the bound state becomes much smaller than the 1D channel, and hence the constituent fermions are required to occupy many transverse states. For our experimental conditions this 3D physics reduces the bound state energy by 35% compared to a purely 1D model. This shift is readily accounted for by renormalizing our chemical potentials (see Methods), and leads to no qualitative change in the physics.

Mora *et al.* studied the three[3] and four[4] body problem in 1D tubes. They found that for weak or moderate interactions the scattering between a pair and a third fermion is predominantly antisymmetric while for very strong interactions ($a_\perp / a_{3D}$) > 1.7 (far beyond the confinement induced resonance) the dominant scattering channel is symmetric. When the scattering between the pair and a third particle is antisymmetric, one encounters an analog of FFLO physics: the three-body wavefunction changes sign when a pair passes a fermion. Such a sign change under the exchange of a boson and a fermion is exotic. The symmetric scattering situation is more common (for example, it is analogous to the repulsive pair-fermion *s*-wave scattering seen on the BEC side of a 3D Feshbach resonance[5]). This change in the symmetry of the scattering qualitatively changes the nature of the many-body state[6], and one expects to see a transition characterized by the correlations of the pair annihilation operator *b(x)*. For sufficiently small $a_\perp / a_{3D}$ one finds FFLO-type correlations[7] $\left\langle b^+(x)b(0)\right\rangle \sim \cos(2\pi n_F x)/|x|^\delta$ where $n_F = n_\uparrow$-$n_\downarrow$ and $\delta$ is an exponent which depends on the interaction strength. Conversely, for sufficiently large $a_\perp/a_{3D}$ one finds[8] $\left\langle b^+(x)b(0)\right\rangle \sim 1/|x|^{\delta'}$. The exact nature of the transition between these behaviours is an area of current research.

Although our current experiments are well in the FFLO regime, the scattering length $a_{3D}$ can be easily tuned by changing the magnetic field, making this transition readily amenable to experimental study.

**Supplementary Notes**

**Figure S1:  The imaging process.  a,** The PCPI imaging setup.  A quarter wave plate (left) converts the linearly polarised probe beam to circular polarisation.  After coherently scattering from the atoms, the then elliptically polarised beam is decomposed by a polariser (right) and imaged with a CCD camera (far right).  **b,** The relevant energy levels and probe beam frequencies.  The first probe S1 is detuned 15 $\Gamma$ from state $|2\rangle$, and the second probe S2 is tuned half way between states $|1\rangle$ and $|2\rangle$.  **c,** Column density images from probes S1 and S2 corresponding to $P = 0.63$.  **d,** Column density profiles of state $|1\rangle$ and $|2\rangle$, and their difference deconvolved from images S1 and S2.

**Figure S2: a**, Scatter plot of best fit temperature and central tube polarisation for the 95 data sets used in Fig. 3. The average temperature of the data set is $T = 175$ +/- 50 nK = 0.15 $T_F$.  **b**, A sample theoretical fit to the axial densities for $P = 0.055$ and **c**, for $P = 0.10$ where temperature is the only free parameter.  Lines: Theoretical fits; black circles: majority; blue diamonds: minority; red squares: difference; dashed lines: phase boundaries.  The fitted temperature for this data is 130 nK.  Inset shows reduced chi-squared as function of $T$.

**Figure S3:** Sample phase boundaries from fitted data sets along the axial direction, $z$, as function of radial position in the tube bundle, $\rho$.  **a,** for central tube polarisation $P = 0.11$ and **b**, for $P = 0.57$.  Blue line: minority; red line: difference.  Inset shows tube polarisation as function of radial position in the tube bundle.

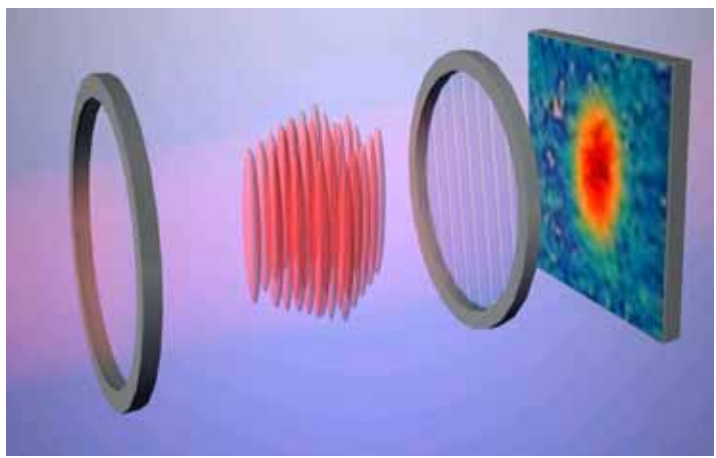

**a**

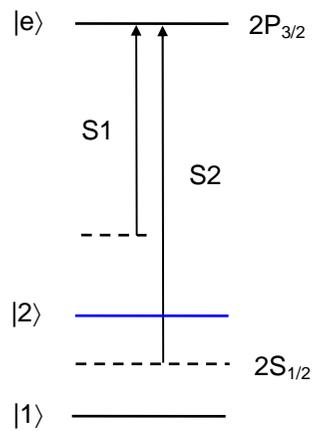

$|e\rangle$ —————— $2P_{3/2}$

S1          S2

$|2\rangle$ ——————

—————— $2S_{1/2}$

$|1\rangle$ ——————

**b**

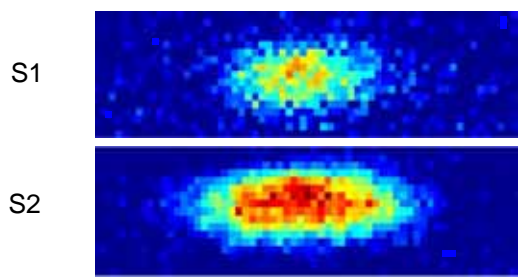

S1

S2

**c**

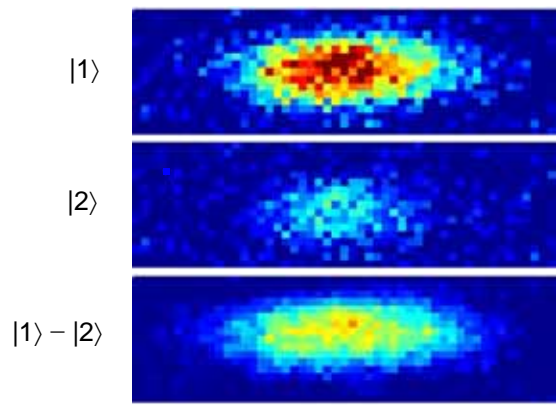

$|1\rangle$

$|2\rangle$

$|1\rangle - |2\rangle$

**d**

Author: Liao et. al.
Figure S1

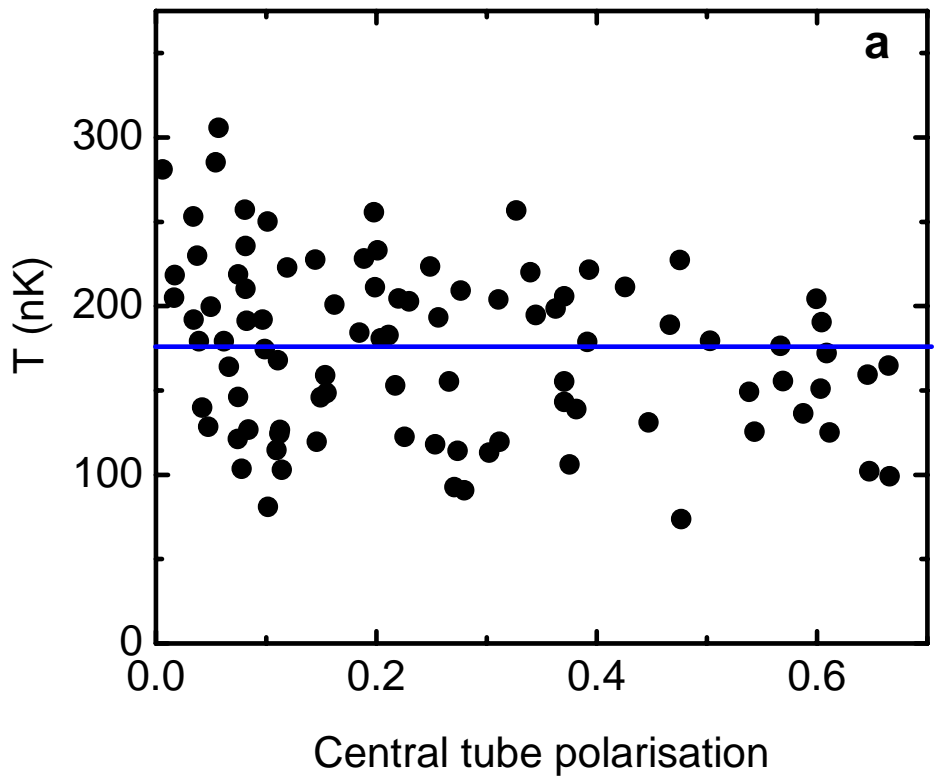

Central tube polarisation

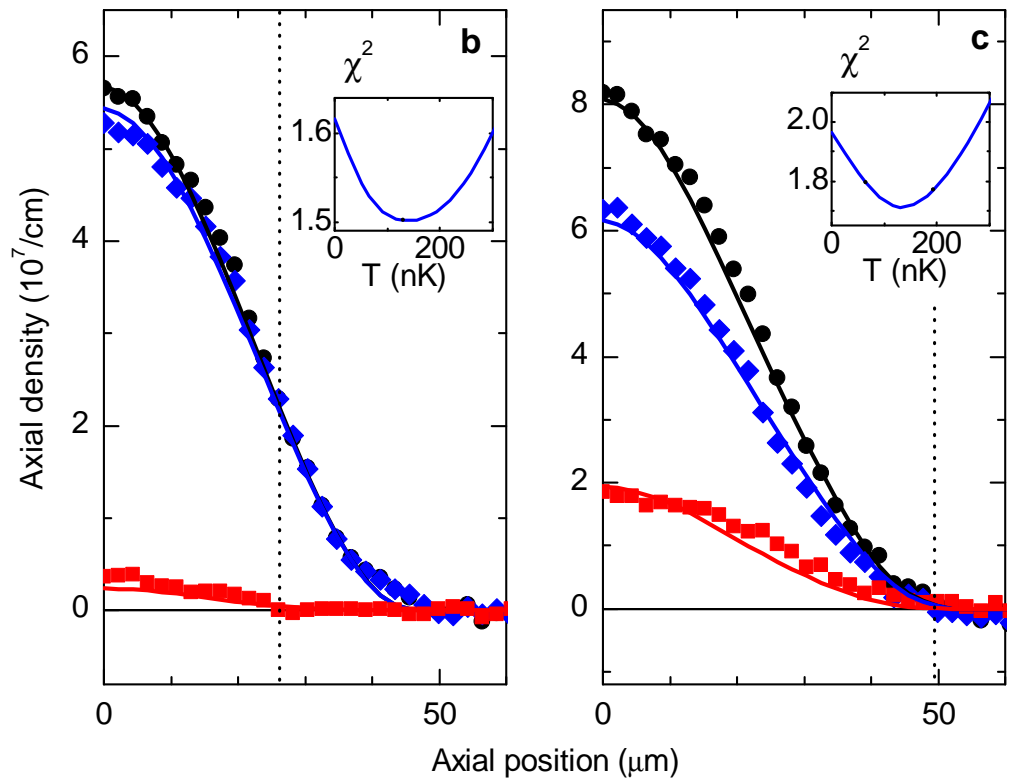

Author: Liao et. al.
Figure S2 bc

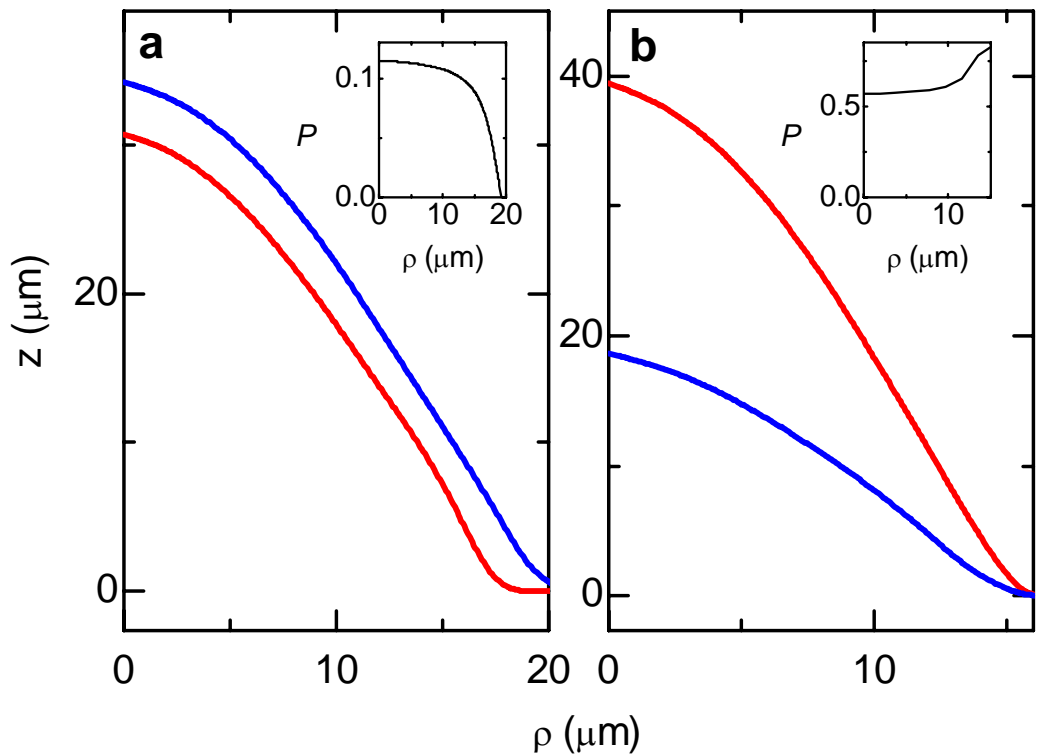